# Chemically Tailored Growth of 2D Semiconductors *via* Hybrid Metal-Organic Chemical Vapor Deposition

*Zhepeng Zhang[1,#], Lauren Hoang[2,#], Marisa Hocking[1], Jenny Hu[3], Gregory Zaborski Jr.[1], Pooja Reddy[1], Johnny Dollard[1], David Goldhaber-Gordon[4,5], Tony F. Heinz[3,5,6], Eric Pop[1,2,7], Andrew J. Mannix[1,5*]*

[1]Department of Materials Science & Engineering, Stanford University, Stanford, CA 94305, USA
[2]Department of Electrical Engineering, Stanford University, Stanford, CA 94305, USA
[3]Department of Applied Physics, Stanford University, Stanford, CA 94305, USA
[4]Department of Physics, Stanford University, Stanford, CA 94305, USA
[5]Stanford Institute for Materials and Energy Sciences, SLAC National Accelerator Laboratory, Menlo Park, CA 94025
[6]Department of Photon Sciences, Stanford University, Stanford, CA 94305, USA
[7]Precourt Institute for Energy, Stanford University, Stanford, CA 94305, USA

*Corresponding author: A.J.M., ajmannix@stanford.edu

#These authors contributed equally to this work (Z.Z., L.H.).

## ABSTRACT

Two-dimensional (2D) semiconducting transition-metal dichalcogenides (TMDCs) are an exciting platform for new excitonic physics and next-generation electronics, creating a strong demand to understand their growth, doping, and heterostructures.

Despite significant progress in solid-source (SS-) and metal-organic chemical vapor deposition (MOCVD), further optimization is necessary to grow highly crystalline 2D TMDCs with controlled doping. Here, we report a hybrid MOCVD growth method that combines liquid-phase metal precursor deposition and vapor-phase organo-chalcogen delivery to leverage the advantages of both MOCVD and SS-CVD. Using our hybrid approach, we demonstrate $WS_2$ growth with tunable morphologies – from separated single-crystal domains to continuous monolayer films – on a variety of substrates, including sapphire, $SiO_2$, and Au. These $WS_2$ films exhibit narrow neutral exciton photoluminescence linewidths down to 33 meV and room-temperature mobility up to 34 – 36 $cm^2V^{-1}s^{-1}$). Through simple modifications to the liquid precursor composition, we demonstrate the growth of V-doped $WS_2$, $Mo_xW_{1-x}S_2$ alloys, and in-plane $WS_2$-$MoS_2$ heterostructures. This work presents an efficient approach for addressing a variety of TMDC synthesis needs on a laboratory scale.

Two-dimensional (2D) semiconducting transition metal dichalcogenides (TMDCs), such as monolayer $MoS_2$, $WS_2$, and $WSe_2$ have emerged as attractive candidates for next-generation electronics due to their atomic-scale thickness, tunable band structure, and excellent electronic properties.[1–3] In the past decade, demonstrations of high-performance 2D TMDC-based transistors, optoelectronics, and logical circuits have escalated demand for the accurately controlled large-area growth of high-quality pure and p-/n-type doped 2D TMDC monolayers.[4–12] Solid source chemical vapor deposition (SS-CVD) has become a popular approach for growing 2D TMDCs in laboratory settings due to its low equipment cost, flexibility, and rapid growth, enabling efficient optimization. By using SS-CVD, a wide range of 2D TMDCs, such as $MoS_2$,[13,14] $WS_2$,[6] V-doped $WSe_2$,[10,11] Fe-doped $MoS_2$,[15] and $Mo_xW_{1-x}S_2$ alloy[16] have been successfully synthesized, and wafer-scale TMDC synthesis and device fabrication have been demonstrated.[5,17]

However, further optimization for SS-CVD growth is both necessary and challenging. For example, solid sources typically exhibit low sublimation rates and poor sublimation stabilities during the material growth process. The solid precursor is challenging to replenish mid-growth, resulting in variable stoichiometry in the reactor over time during each growth run. Small variations in source amount and position modify the uniformity of the growth. These factors limit the tolerance and controllability of SS-CVD.[18,19] Moreover, though a specific SS-CVD strategy normally works well for an individual TMDC system, a universal method for multiple-material synthesis remains underdeveloped. Even though the situation has been improved by new source supply strategies[20,21] and adding promoters,[22–24] the design of state-of-the-art SS-CVD growth setups has also become increasingly complex – and correspondingly, less accessible – for most laboratory research.

On the other hand, metal-organic chemical vapor deposition (MOCVD) has shown good reproducibility and large-area uniformity in 2D TMDC growth[25] at relatively low reaction temperatures (150–320 °C)[26–28] and under accurate precursor control due to the use of vapor phase metal-organic metal (M-Organic) and hydride or organic chalcogen (X-Organic) precursors.[25] However, to reduce carbon impurity incorporation, MOCVD often uses low precursor concentrations, resulting in slow growth rates of 2D TMDCs. Moreover, each novel dopant metal-organic source requires a separate precursor supply line in the MOCVD system to avoid cross contamination, which increases the system cost and complexity and hinders the exploration of substitutional doping. Despite the development of new MOCVD strategies to enlarge the domain size,[29] enable epitaxy,[30] and reduce the growth temperature,[26,27] more accessible and efficient MOCVD growth and doping methods are still needed.

Here, we report a hybrid MOCVD (Hy-MOCVD) growth method, which delivers metal precursors and growth promoters from the solution phase and metal-organic chalcogen precursors from the vapor phase, to combine the advantages of both MOCVD and SS-CVD and realize efficient growth of multiple types of 2D TMDCs. Aqueous Hy-MOCVD precursor delivery by both spin-coating and dip-coating produce $WS_2$ monolayers with good controllability and uniformity. Hy-MOCVD grown $WS_2$ exhibits typical domain sizes of tens of micrometers, good optical quality with room temperature neutral exciton peak width down to 33 meV, and good electronic performance with electron mobility up to $34 - 36$ $cm^2V^{-1}s^{-1}$, and transistor on/off ratio of $>10^7$. Hy-MOCVD also enables the growth of $WS_2$ on diverse substrates, such as c-plane and m-plane sapphire, $Si/SiO_2$, and sapphire/Au. To illustrate the versatility of our Hy-MOCVD approach, we also demonstrate the facile growth of V-doped $WS_2$, $Mo_xW_{1-x}S_2$ alloys, and $WS_2$-$MoS_2$ heterostructures without any modifications to the growth hardware.

**RESULTS AND DISCUSSIONS**

In **Figure 1**, we compare the concepts and strengths of SS-CVD, MOCVD and Hy-MOCVD. The Hy-MOCVD method employs both organic chalcogen precursors (X-organic) used in MOCVD and inorganic transition metal precursors (M-inorganic) used in SS-CVD. As in MOCVD, X-organic was introduced into the Hy-MOCVD chamber in the vapor phase *via* a bubbler and a mass flow controller (see **Figure S1** for the set-up schematic of Hy-MOCVD). This ensures a stable chalcogen concentration throughout the entire growth process, which is necessary for stoichiometrically controlled growth. Precise combinations of the primary transition metal element(s), substitutional dopants, and any growth promoter species are more challenging to deliver due to their lower vapor pressure, yet these are also critical to the outcome of the growth process.[23,26] To overcome the uncontrolled flux of solid-source powders and the expense of metal-organic precursor delivery, M-inorganic precursor with growth promoter KOH were deposited onto the growth substrate by aqueous solution coating before the Hy-MOCVD growth. This localized transition metal supply ensures a high concentration of reactive M species on the wafer surface during growth. Moreover, by mixing M-inorganic and KOH with other dopant sources,[8,10,31,32] Hy-MOCVD can be used for the growth of doped TMDCs and TMDC alloys with extreme precision *via* dilution.[9,32] Summarizing these advantages, Hy-MOCVD combines the precise control over chalcogen stoichiometry found in MOCVD with the versatility and efficiency in switching or mixing transition metals and growth promoters offered by SS-CVD. In the following sections, we will demonstrate these advantages by using Hy-MOCVD to grow $WS_2$ and incorporate dopants, alloys, and heterostructures.

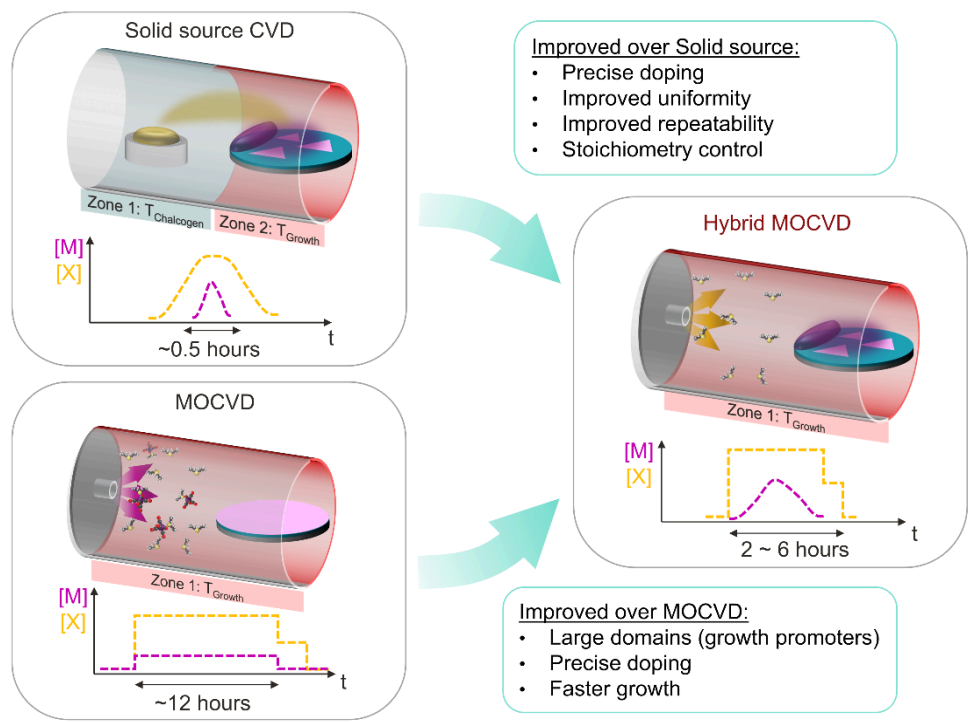

**Figure 1. Principles of Hy-MOCVD.** Representative schematics of the growth setups and precursor supply time profiles for conventional SS-CVD, MOCVD and Hy-MOCVD. The *y*-axis in the time profiles stands for the active concentrations of the transition metal (M) and chalcogen (X) species.

In the Hy-MOCVD growth of $WS_2$, diethyl sulfide (DES, $(CH_3CH_2)_2S$) and ammonium metatungstate hydrate (AMT, $(NH_4)_6H_2W_{12}O_{40} \cdot xH_2O$) were used as the X-organic and M-inorganic precursors, respectively. Delivery of the metal solution to the substrate is flexible, and we explored two paths in this work: spin-coating and dip-coating (**Figure 2a**). In spin-coating delivery, the starting solution of AMT and KOH in deionized (DI) water was spin coated onto a UV-ozone treated wafer, and the water was removed by heating at 80 °C in air. The coated wafer was then transferred to the tube furnace MOCVD system and annealed in a DES vapor environment (0.05-0.12 sccm) at 775 °C for 2-6 h to conduct the growth. Photographs of a typical $WS_2$ on *c*-plane sapphire wafer after the growth show uniform color across the wafer (**Figure 2b**). Optical microscope images show homogeneous coverage of $WS_2$ triangular domains, typically ~20 μm in width, with sharp and straight edges (**Figure 2c**). Atomic force microscopy (AFM) shows the monolayer thickness and clean surface of Hy-MOCVD grown $WS_2$ (**Figure 2d**). A typical photoluminescence (PL) spectrum (**Figure 2e**) shows a strong and narrow neutral exciton peak (A) at 1.99 eV with full width half maximum (FWHM) of 36 meV, indicating the good quality of Hy-MOCVD grown $WS_2$ .[33,34] The lower-energy shoulder peak is attributed to the negatively charged exciton (A-), consistent with the n-type electronic transport characteristics observed in Hy-MOCVD $WS_2$ monolayers discussed in a later section.

In dip-coating delivery, the *c*-plane sapphire wafer edges were dipped into aqueous solution of AMT and KOH. As for spin-coating, the wafer was then dried in air at 80 ºC, and annealed in DES. During the growth process, reactive species diffuse from the highly concentrated AMT+KOH sources at the sample edges, triggering the growth of $WS_2$ on the uncoated center area of the wafer. Typical photos of the wafer show deeper color on the dip-coated edges and uniform light yellow-green in the center of the wafer (**Figure 2f**). An optical micrograph taken from the center of the wafer shows a continuous $WS_2$ film with small multilayer islands (**Figure 2g**). AFM images acquired around a multilayer island show well-defined single-layer-height steps of the bilayer island and clear atomic steps and terraces of the *c*-plane sapphire substrate visible through the monolayer, indicating the clean surface of $WS_2$ film (**Figure 2h**). PL spectra collected from continuous monolayer regions of these samples typically show A exciton peaks centered at 2.00 eV (33 meV FWHM), consistent with a good-quality monolayer film (**Figure 2i**).[33] We have found that both spin coating and dip coating yield good-quality and consistent growths.

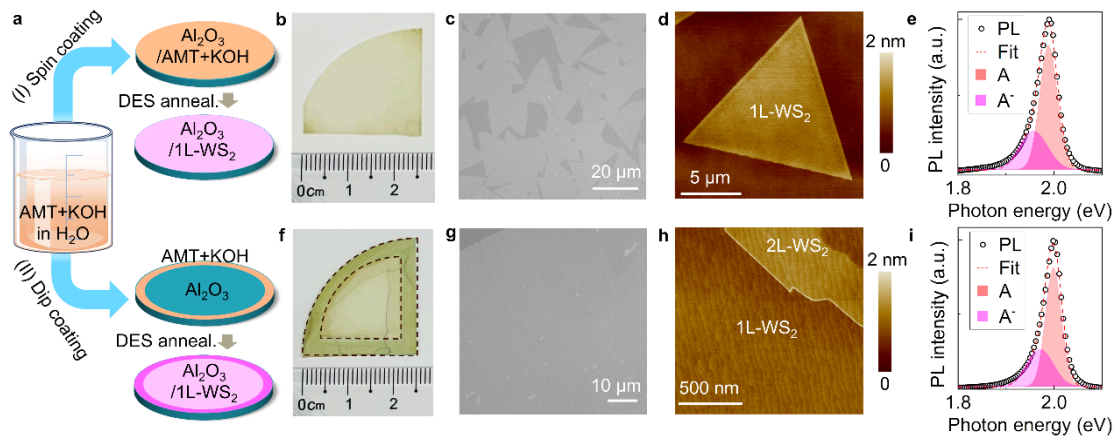

**Figure 2. Hybrid MOCVD processes. (a)** Schematics of the two paths to hybrid MOCVD: (I) spin coating and (II) dip coating. **(b)** Typical photo of sapphire/WS$_2$ wafer, **(c)** contrast-enhanced optical image, **(d)** AFM image and **(e)** PL spectrum of monolayer WS$_2$ growth with spin coating (path I). **(f-i)** Same images for dip coating (path II). Dashed lines in **(f)** highlight the dip-coated area on the edges of sapphire wafer.

Growth producing a well-defined compositional gradient can be valuable for exploratory synthesis. Dip-coating Hy-MOCVD can exploit the vapor-phase transport gradient to grow WS$_2$ with different morphologies and high compatibility with different substrates. **Figure 3a** is a photo of the *c*-plane sapphire wafer after Hy-MOCVD growth with only one edge coated with AMT+KOH solution. The WS$_2$ coverage changes with increasing distance from the dip-coating boundary (**Figure 3b-e**), with a typical profile given by **Figure 3f** (extracted from binary thresholding of microscope images; coverage over 100% indicates multilayer islands over a continuous monolayer film). At higher magnification within these regions, we observe that WS$_2$ grew as a continuous film with a high density of multilayer islands in the area close to the dip-coating boundary (**Figure 3g**). This converts to a continuous monolayer with low density of multilayer islands in the center of the wafer (**Figure 3h**) and finally becomes isolated domains on the far end (**Figure 3i**). The high coverage region (>70%) of predominantly monolayer WS$_2$ extends to approximately 1 cm away from the dip-coating metal source region, which is typical of samples grown in this way. Ozone etching reveals the grain boundaries[35] within the continuous $WS_2$ regions (**Figure S2**), and we observe that the average $WS_2$ domain size varies from 3 to 30 μm with increasing distance from the dip-coating boundary.

Our results suggest that optimization of the dip-coating strategy enables Hy-MOCVD continuous films over large areas. **Figure 3j** displays $WS_2$ growth over a 2" *c*-plane sapphire wafer by dip coating the wafer edge and placing two AMT+KOH dip-coated W foil strips on top of the substrate to provide enough W-species for the $WS_2$ growth near the wafer center during the Hy-MOCVD growth.

Growth on multiple substrates is important for laboratory-scale optimization and integration of TMDCs. Hy-MOCVD growth of $WS_2$ on annealed *a*-plane sapphire substrates with 1° miscut angle towards the *c*-plane (**Figure 3k**) resulted in $WS_2$ ribbons oriented along the substrate <1-100> terrace edge direction (see **Figure S3** for the AFM images). This morphology is consistent with previous observations of epitaxial growth of $WS_2$ on *a*-plane sapphire *via* SS-CVD.[7,36] Polarization-resolved second-harmonic generation (SHG) reveals that the Hy-MOCVD grown $WS_2$ ribbons exhibit predominantly two sets of epitaxial lattice orientations, with the $WS_2$ armchair directions oriented parallel to either the <1-100> or <0001> directions of the *a*-plane sapphire (**Figure S4**). This observation suggests that Hy-MOCVD can realize van der Waals epitaxial growth of 2D TMDCs and enhance the understanding of how precursors and alkali metal-based growth promoters[37] modify epitaxy. Additionally, Hy-MOCVD is compatible with the growth of $WS_2$ on standard thermally oxidized $Si/SiO_2$ substrates and on Au thin films deposited on *c*-plane sapphire substrates (**Figure 3l,m**). Notably, the Raman out-of-plane mode ($A_1'$) of $WS_2$ on Au exhibits a redshift of ~7 cm$^{-1}$, shifting the peak center to 410 cm$^{-1}$, while the in-plane mode (E') remains unaltered at ~354 cm$^{-1}$ compared to WS$_2$ grown on SiO$_2$ (**Figure 3n**). This observation aligns with the reported A$_1$' mode downshifting in exfoliated WS$_2$ monolayer on Au, and suggests the strong interaction between monolayer WS$_2$ and Au.[38] PL of WS$_2$ grown on Si/SiO$_2$ confirms its high quality, whereas the quenched PL for WS$_2$ grown on Au indicates non-radiative transitions dominated recombination of excitons in Au/WS$_2$ stack (**Figure 3o**). Furthermore, Hy-MOCVD WS$_2$ on different substrates exhibited an absence of Raman peaks within the 1300-1600 cm$^{-1}$ range (See **Figure S5** for the Raman spectra), indicating that the films are free of amorphous carbon.

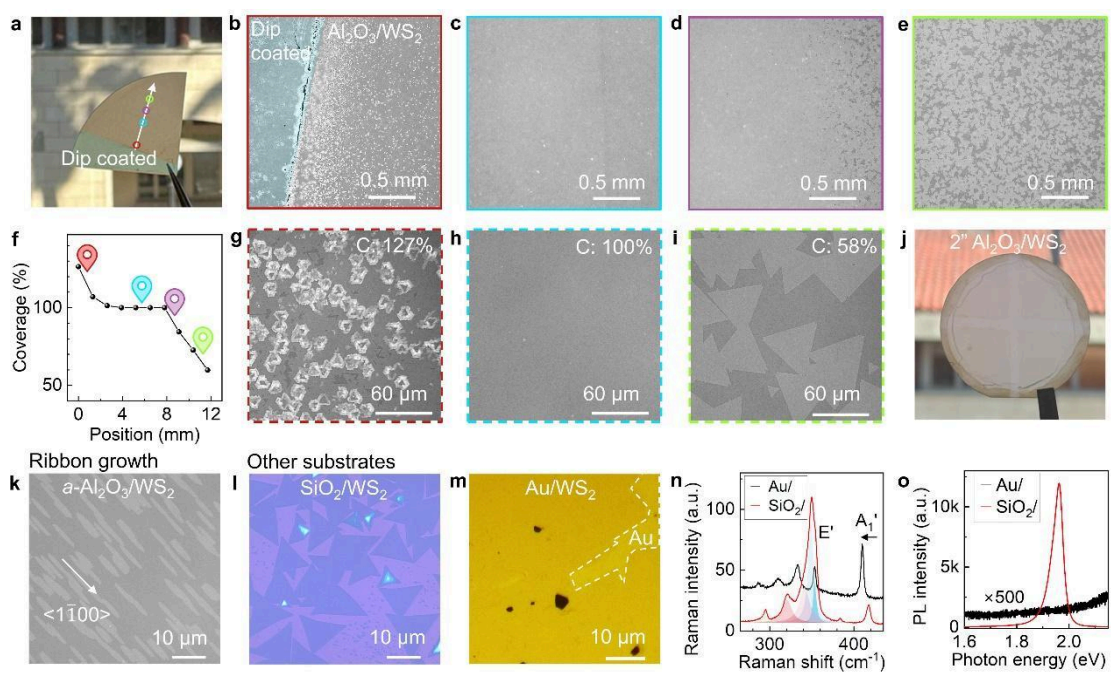

**Figure 3. Morphology control and compatibility with other substrates. (a)** Photo of Hy-MOCVD grown sapphire/WS$_2$ wafer with single edge dip coated by precursor solution. **(b-e)** Contrast-enhanced optical images of sapphire/WS$_2$ taken from the locations highlighted by colored circles in **(a)**. **(f)** WS$_2$ coverage versus position along the arrow in **(a)**. The positions of **(b-e)** are highlighted with corresponding colors. **(g-i)** Contrast-enhanced zoom-in optical images of multilayer, continuous monolayer, and non-continuous monolayer regions. $C$ stands for the coverage extracted from the corresponding image. **(j)** Photo of Hy-MOCVD grown WS$_2$ on a 2" sapphire wafer

*via* optimized dip-coating path. **(k)** Optical image of Hy-MOCVD grown WS$_2$ ribbons on annealed *a*-plane sapphire with 1° miscut angle towards c-plane. **(l)** Optical image of Hy-MOCVD WS$_2$ grown on Si/SiO$_2$ substrate. **(m)** Optical image of Hy-MOCVD grown WS$_2$ grown on sapphire/Au substrate. **(n,o)** Raman and PL spectra of WS$_2$ grown on SiO$_2$ and Au substrates, respectively.

To evaluate the electronic properties of Hy-MOCVD grown WS$_2$, we fabricated back-gated field-effect transistors (FETs) through two processes: by transferring the Hy-MOCVD monolayer WS$_2$ from sapphire onto SiO$_2$ (100 nm) on highly-doped p$^{++}$ Si, and by using as-grown Hy-MOCVD monolayer WS$_2$ directly on similar substrates. FET channel regions (100 nm to 1 μm) were defined by electron-beam lithography on WS$_2$ triangular domains and contacted with Ni/Au electrodes to achieve transfer length method (TLM) structures (**Figure 4a,b**).[39] Measured drain current vs back-gate voltage ($I_D$ vs $V_{GS}$) characteristics of such WS$_2$ FETs exhibit consistent n-type behavior across 10-17 devices for each channel length, illustrating the uniformity of Hy-MOCVD grown WS$_2$ (**Figure 4c,d**).

The devices with transferred WS$_2$ exhibit maximum electron mobility between 24 – 33 cm$^2$V$^{-1}$s$^{-1}$ (this value is given as a range of two numbers, extracted from the forward and backward sweeps, due to the observed clockwise hysteresis), with average value between 13 – 18 cm$^2$V$^{-1}$s$^{-1}$ and median value between 13 – 19 cm$^2$V$^{-1}$s$^{-1}$. We see a notable average $I_{max}/I_{min}$ ratio of 10$^7$ (**Figure 4e,f**). The shortest devices with a 100 nm channel length have a good on-state current density, reaching a maximum value of 88 μA/μm and an average of 65 μA/μm at $V_{DS}$ = 1 V (See **Figure S6a** for $I_D$ versus channel length plot). These metrics surpass those of most SS-CVD and MOCVD-grown monolayer WS$_2$-based FETs with similar configurations, indicating the good quality of Hy-MOCVD WS$_2$ (**Table S1** for device performance comparison). The contact resistance can lead to errors in the field-effect mobility estimate, especially in shorter channel length devices (**Figure S6c** shows field-effect mobility versus channel length). The device performance can potentially be improved by incorporating lower resistance contacts and high-κ dielectric layers.[40–43] FET devices fabricated from Hy-MOCVD $WS_2$ grown directly on the Si/$SiO_2$ substrate exhibit improved field-effect mobility with maximum between $34 - 36$ cm$^2$V$^{-1}$s$^{-1}$, an average value of $19 - 21$ cm$^2$V$^{-1}$s$^{-1}$ and a median value of $20 - 22$ cm$^2$V$^{-1}$s$^{-1}$ (**Figure 4e**), and less $I_D$ hysteresis for forward-to-backward $V_{DS}$ sweeps (**Figure 4c,d** and **Figure S6c-h**). This suggests that the performance of Hy-MOCVD grown on sapphire substrates is limited by either transfer-induced damage (See broadened Raman and PL peaks of $WS_2$ after the transfer in **Figure S7**) or a difference in crystal quality versus growth on Si/$SiO_2$ substrates.

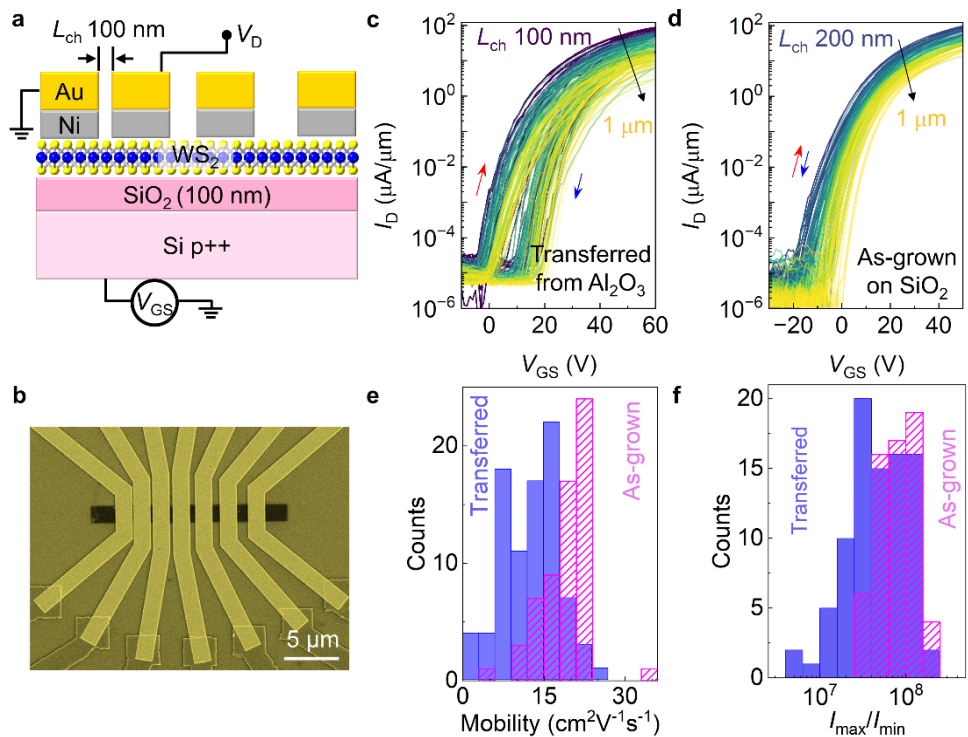

**Figure 4. Electrical characteristics of monolayer $WS_2$ grown by hybrid MOCVD.**
(**a**) Schematic of a back-gated transistor based on Hy-MOCVD $WS_2$. (**b**) False color SEM image of $WS_2$-based TLM device. (**c**) Measured $I_D$ vs $V_{GS}$ curves for FETs of transferred Hy-MOCVD $WS_2$ with designed channel length $L_{ch}$ of 100, 200, 300, 500, 700 and 1000 nm, from purple to yellow at $V_{DS}$ = 1 V. Red and blue arrows represent the forward and backward $V_{GS}$ sweeping directions, respectively. (**d**) Measured $I_D$ vs $V_{GS}$ curves for FETs of as-grown Hy-MOCVD WS$_2$ with designed channel length $L_{ch}$ of 200, 300, 500, 700 and 1000 nm (From blue to yellow). Red and blue arrows represent the forward and backward $V_{GS}$ sweeping directions, respectively. Histograms of measured (**e**) field-effect mobility and (**f**) $I_{max}/I_{min}$ for FETs of transferred and as-grown Hy-MOCVD WS$_2$ (Extracted from forward $V_{GS}$ sweeps).

Directly incorporating dopants into TMDCs and growing TMDC alloys and heterostructures from synthesis have sparked substantial interest. Hy-MOCVD enables convenient adjustment of the TMDC metal composition based on the precise addition of various water-soluble transition metal sources into the precursor solution (**Figure 5a**). V-doped WS$_2$ monolayers with nominal doping from 0.3 to 24% (V/(V+W) atom mole ratio in the precursor solution) were grown on Si/SiO$_2$ substrates (**Figure 5b,c**) by adding sodium metavanadate (NaVO$_3$) into the AMT+KOH precursor solution. The emergence of a new Raman mode at around 213 cm$^{-1}$ in nominal 24% V-doped WS$_2$, and the decrease of the 2LA(M)+ E' peak intensity with the increase of nominal doping ratio, are consistent with previous V-WS$_2$ literature (see **Figure 5d** and **Figure S8** for nominal doping ratio dependence of 2LA(M)+ E' peak intensity).[10,44] The characteristic peak at 213 cm$^{-1}$ can be assigned to the multi-phonon mode of E"(M)-TA(M), suggesting that V is substitutionally incorporated into WS$_2$.[44] Transistors fabricated using the 3% V-doped WS$_2$ exhibit a threshold voltage shift of +23 V compared with undoped WS$_2$ devices (**Figure 5e**), on the 100 nm SiO$_2$ back-gate insulators. This is consistent with the expected p-type doping from substitutional V acceptors in the TMDC monolayer.[10,32] Additional optimizations of doping concentration and FET metal contacts are needed to achieve hole current.

Hy-MOCVD similarly enables alloy and heterostructure growth. We grew $Mo_xW_{1-x}S_2$ alloys exhibiting an in-plane heterostructure with a core and a shell of difference alloy compositions during a single-step dip-coating Hy-MOCVD growth (**Figure 5f,g**) by mixing ammonium molybdate $((NH_4)_6Mo_7O_2 \cdot 4H_2O)$ into the AMT+KOH solution. The Mo/W molar ratio of the precursor solution influenced the $Mo_xW_{1-x}S_2$ alloy core-shell dimension and alloy compositions, as illustrated in **Figure 5h,i**. For example, a 2:1 Mo/W ratio yielded a $MoS_2$ core with a $WS_2$-like alloy shell (*i.e.*, an alloy closer in Raman signature to the signature of pure $WS_2$), whereas the decrease to a 1:8 Mo/W ratio resulted in a $MoS_2$-like core with a $WS_2$ shell. The core-shell structure evidently results from differences in vapor-phase or on-surface transport kinetics for the W and Mo precursors.[45,46]

In contrast, two sequential Hy-MOCVD growths of W followed by Mo precursors resulted in $WS_2$-$MoS_2$ in-plane heterostructures (**Figure 5j,k**). Raman spectra collected from two sides of the $WS_2$-$MoS_2$ heterostructure show distinct $MoS_2$ and $WS_2$ peaks without significant alloying (**Figure 5l**), and a Raman spectrum line scan shows a distinct interface between $WS_2$ and $MoS_2$ (**Figure 5m**).

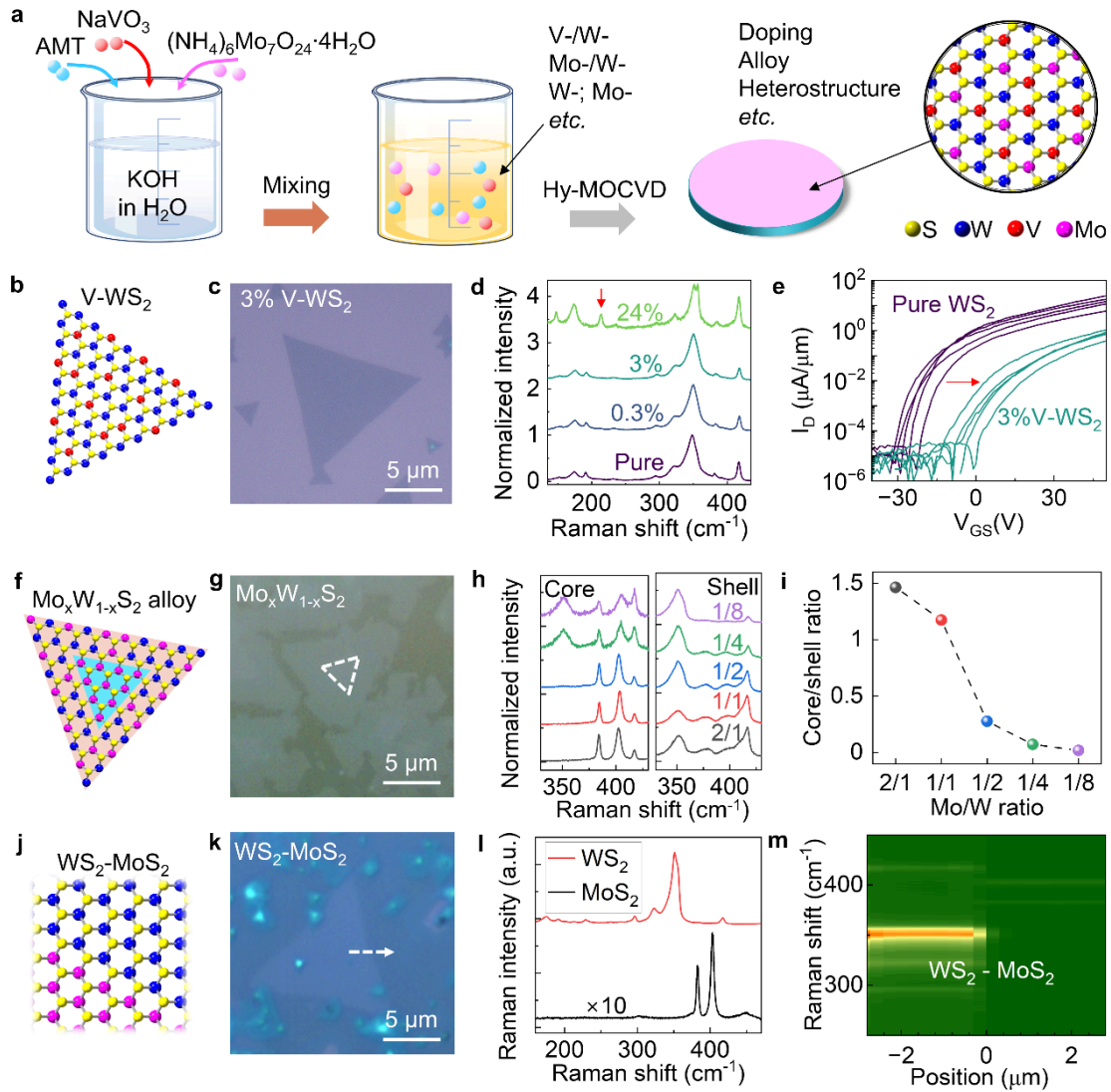

**Figure 5. Transition metal engineering of monolayer WS₂ using hybrid MOCVD.** (**a**) Schematic of transition metal engineering of monolayer WS₂ using hybrid MOCVD. (**b**) Lattice schematic of V-doped WS₂. (**c**) Optical image of as-grown Hy-MOCVD V-doped WS₂ on Si/SiO₂ substrate with a nominal doping concentration of 3%. (**d**) Typical Raman spectra of V-doped WS₂ with different nominal doping ratios of 0, 0.3, 3 and 24%. (**e**) Measured $I_D$ vs $V_{GS}$ curves for monolayer undoped WS₂ and V-doped WS₂ FET devices with channel length of 500 nm and $V_{DS}$ = 1 V. (**f**) Lattice schematic of in-plane MoS₂-Mo$_x$W$_{1-x}$S₂ heterostructure with a MoS₂ core and a Mo$_x$W$_{1-x}$S₂ alloy shell. (**g**) Optical image of Hy-MOCVD grown in-plane MoS₂-Mo$_x$W$_{1-x}$S₂ heterostructure on sapphire substrate. A MoS₂ core is circled with a white dashed line. (**h**) Typical Raman spectra of Mo$_x$W$_{1-x}$S₂ alloy core (left) and shell (right) grown with different Mo/W mole ratios in the starting solution of Hy-MOCVD. (**i**) Core/shell width ratio versus Mo/W mole ratio of starting solution. (**j**) Lattice schematic of Hy-MOCVD grown WS₂-MoS₂ in-plane heterostructure. (**k**) Optical image of Hy-MOCVD grown WS₂-MoS₂ in-plane heterostructure. (**l**) Typical

Raman spectra collected from the two sides of $WS_2$-$MoS_2$ in-plane heterostructure. **(m)** Raman spectra line scan along the arrow in **(k)**.

## CONCLUSIONS

In summary, we have demonstrated Hy-MOCVD enables the synthesis of good-quality monolayer $WS_2$ with simplified growth equipment and can be easily extended to produce V-doped $WS_2$, $Mo_xW_{1-x}S_2$ alloys, and $MoS_2$-$WS_2$ heterostructures. This method provides an effective strategy for rapidly synthesizing TMDC monolayers with diverse transition metal dopants, alloy elements, and heterostructures, offering a versatile platform for exploring synthesis to realize new electronic, optical, and magnetic properties in TMDC monolayers and heterostructures.

## EXPERIMENTAL METHODS

***Material growth and transfer.*** *Hy-MOCVD commenced with the preparation of an initial aqueous solution comprising transition metal precursors and promoters. In the case of pure $WS_2$ growth, 0.6 g of AMT and 0.05-0.1g of KOH were dissolved in 30 ml DI water. For V-doped $WS_2$ growth, around 90 mg of $NaVO_3$ was introduced into the 30 ml AMT+KOH solution to achieve 24% V/(V+W) atom mole ratio in the solution. Ultrasonication was employed to facilitate the dissolution of $NaVO_3$. $NaVO_3$ was not fully dissolved, and the cloudy solution was used for growing 24% V-$WS_2$ sample. The cloudy solution was diluted multiple times to get 3% and 0.3% V/(V+W) atom mole ratio solutions. In these low V/(V+W) ratio solutions, $NaVO_3$ appeared to be fully dissolved. For $Mo_xW_{1-x}S_2$ Alloy growth, AMT + KOH (0.6 g + 0.05 g in 30 ml DI water) and ammonium molybdate + KOH (0.43 g + 0.2 g in 30 ml DI water) solutions were made separately and mixed with different volume ratios from 2/1 to 1/8. For the growth of $WS_2$-$MoS_2$ heterostructures, two-step dip-coating Hy-MOCVD were used to grow $WS_2$ and $MoS_2$ sequentially. In the dip-coating path of Hy-MOCVD, the aqueous solution was dip-coated onto one or all edges of ozone-treated sapphire substrates,* *followed by N₂ blow drying. For the dip-coating Hy-MOCVD growth on a 2" c-plane*
*sapphire wafer, in addition to coating the wafer edge, two initial solution coated W*
*foil strips were placed on the top of the wafer, forming a cross and siting at its center.*
*In the spin-coating path of Hy-MOCVD, 0.25 ml of 10-16 times diluted initial solution*
*was spin coated onto ozone-treated sapphire and Si/SiO₂ substrates at 1000 rpm for 1*
*min, employing 0.25 ml of the solutions. When growing on Si/SiO₂ and sapphire/Au,*
*no ozone was applied before dip coating. For the growth of WS₂ on a-plane sapphire*
*(Hefei Crystal Technical Material Co. Ltd, a-plane off c-plane 1.0°±0.1°), the wafer*
*was annealed in a muffle furnace at 1200 °C for 12 hours in an ambient air*
*environment. The c-plane sapphire wafers (Valley Design Corp., 28362-1) used in this*
*paper were not annealed. The solution-coated substrates were baked on a hotplate at*
*80 °C for 1 min and quickly loaded into a MOCVD tube furnace. The tube was*
*evacuated to <0.5 Torr and filled with a flowing mixture of 1600 sccm Ar and 10 sccm*
*H₂. The furnace temperature was ramped to 725-775 °C over 30 min. Subsequently,*
*the H₂ flow was adjusted to 1 sccm, and 0.05-0.12 sccm of DES was introduced into*
*the tube furnace. The substrates underwent annealing in this environment for 2-6*
*hours to complete growth. Post-growth, the DES flow was reduced to 0.025-0.1 sccm*
*and the furnace heating was discontinued. DES flow was closed when the furnace*
*naturally cooled down to 300 °C and substrates were unloaded at room temperature.*

*WS₂ grown on sapphire substrates was transferred onto Si/SiO₂ substrates using a*
*polymethyl methacrylate (PMMA)-assisted transfer method. The samples were spin*
*coated with PMMA and dried on a hotplate at 100 °C for 3 min. WS₂/PMMA was*
*delaminated from the sapphire substrate by gradually dipping the substrate into DI*
*water (the substrate was in an upward-facing position and angled at 30-60° relative*
*to the water surface) and transferred onto the target substrate with SiO₂ (100 nm) on*
*Si, followed by drying on a hotplate at 100 °C for 5 min. The PMMA layer was*
*removed by soaking in acetone at 60 °C for 15 min.*

***Device fabrication and analysis.*** *For transferred devices shown in **Figure 4c**,*
*monolayer WS₂ was grown on sapphire and transferred off using a PMMA-based*
*transfer (as described above) onto 100 nm SiO₂ on Si. For the devices fabricated on*

*the Si/SiO₂ growth substrate, shown in **Figure 4d,** monolayer WS₂ was directly grown on SiO₂ (100 nm) on p⁺⁺ Si (≤0.005 Ω-cm) that also served as the back-gate. Alignment marks were first patterned on the direct-grown sample such that discrete WS₂ crystals could be identified. Electron-beam lithography was employed for each lithography step. Large probing pads (SiO₂/Ti/Pt 10/2/20 nm) were first patterned and deposited by electron-beam evaporation via lift-off. SiO₂ was used in the probing pad to limit pad-to-substrate leakage. XeF₂ was used for channel definition, and the contact region was patterned for lift-off. 15/30 nm Ni/Au contacts were electron-beam evaporated at ~10⁻⁸ Torr and a rate of 0.5 Å/s. 20/35 nm Ni/Au contacts were deposited for the non-transferred devices. The fabricated transistors were measured in a Janis ST-100 probe station at ~10⁻⁴ Torr vacuum using a Keithley 4200 semiconductor parameter analyzer.*

*For undoped and the V-doped WS₂ devices shown in **Figure 5e**, the starting WS₂ and V-WS WS₂ were grown on 100 nm SiO₂ on Si. Alignment marks were patterned to identify monolayer regions on both samples. Metal pads and channels were defined by e-beam lithography as described above. For both the doped and undoped WS₂, Ru/Au (5/50 nm) were deposited via e-beam evaporation to investigate potential p-type transport from V-doped WS₂, based on previous reports of good p-type performance from Ru contacts[42]. The devices were measured under vacuum as described above.*

*Threshold voltage was extracted at a constant current of 10 nA/μm.[47] The field-effect mobility, $\mu_e$ = max($g_m$)/[$C_{ox}V_{DS}(W_{ch}/L_{ch})$], was estimated using the maximum transconductance of forward and backward $V_{GS}$ sweeps, $g_m$=d$I_D$/d$V_{GS}$, and the gate insulator capacitance per unit area is $C_{ox} = \varepsilon_0\kappa_{ox}/t_{ox}$. The SiO₂ gate oxide thickness $t_{ox}$ = 100 nm, oxide relative permittivity $\kappa_{ox}$ = 3.9, $\varepsilon_0$ is the vacuum permittivity, and $V_{DS}$ = 1 V. $W_{ch}$ and $L_{ch}$ are channel width and length, respectively. The designed $W_{ch}$ was 2.0 μm. The final $W_{ch}$ was measured via SEM to be 1.6 μm for FETs of transferred WS₂ and 2.0 μm for FETs of as-grown WS₂. The final $L_{ch}$ in the FETs of transferred WS₂ were measured via SEM to be 72, 175, 261, 461, 650 and 973 nm, corresponding to the designed $L_{ch}$ values of 100, 200, 300, 500, 700, and 1000 nm, respectively. The* *final $L_{ch}$ in the FETs of as-grown $WS_2$ were measured via SEM to be 173, 275, 477, 681 and 993, corresponding to the designed $L_{ch}$ values of 200, 300, 500, 700, and 1000 nm, respectively. The mobilities of the FETs were corrected with these measured $W_{ch}$ and $L_{ch}$ values.*

***Material characterizations.*** *AFM imaging was conducted utilizing a Bruker ICON AFM using the ScanAsyst topography imaging mode with NSC19/Al-BS tip. Raman and PL spectra were acquired at room temperature with 532 nm laser excitation using HORIBA Scientific LabRAM HR Evolution confocal microscope. Optical microscope imaging was performed using an Olympus BX-51 microscope in epi-reflection geometry. The optical microscope contrast for images in **Figure 2 (c,g)**, **Figure 3 (b-e**, **Figure 3 (g-i)**, and **Figure 3k** were enhanced in the following way: after acquisition, we converted the color images to greyscale and increased the contrast and brightness to improve visibility of the $WS_2$ on the transparent sapphire wafer. SHG was performed using a femtosecond laser (NKT Origami Onefive 10, 1030 nm, <200 fs) at room temperature. A 40x objective lens was used to excite the sample with an average power of 5 – 10 mW, and signal was collected in a reflective geometry by an EMCCD (Andor iXon Ultra) with an integration time of 100 ms at each polarization angle.*

## ASSOCIATED CONTENT

The Supporting Information is available free of charge.

Schematic of Hy-MOCVD setup, continuous $WS_2$ monolayer domain size extraction, AFM of $WS_2$ grown on *a*-plane sapphire, SHG characterizations of $WS_2$ ribbons grown on *a*-plane sapphire, wide range Raman spectra of $WS_2$ grown on sapphire and $SiO_2$, $L_{ch}$ vs $I_D$, mobility, hysteresis, and typical $I_D$ vs $V_{GS}$ hysteresis, PL and Raman comparison between as-grown and transferred $WS_2$, Raman 2LA+ E' peak intensity vs nominal doping concentration of V-doped $WS_2$.

**AUTHOR INFORMATION**

**Corresponding Author**

**Andrew J. Mannix**—Department of Materials Science & Engineering, Stanford University, Stanford, CA 94305, USA; Stanford Institute for Materials and Energy Sciences, SLAC National Accelerator Laboratory, Menlo Park, CA 94025

Email: ajmannix@stanford.edu

**Author Contributions**

 Z.Z. and L.H. contributed equally to this work. Z.Z. and L.H. developed the growth recipe under the supervision of A.J.M. and E.P.. Z.Z. performed the material growth and characterizations. L.H. fabricated the devices and conducted the device measurements and analysis under the supervision of A.J.M. and E.P.. M.H. developed the sapphire annealing recipe with J.D. under the supervision of A.J.M.. M.H. performed SHG measurements with the help of J.H. under the supervision of T.F.H.. P.R. and G.Z.Jr. built the MOCVD system under the supervision of A.J.M.. Z.Z. and A.J.M. wrote the paper with input from L.H.. All authors participated in discussions, reviewed, and approved the manuscript.

**ACKNOWLEDGMENTS**

This work was supported primarily by the U.S. Department of Energy (DOE), Office of Science, Office of Basic Energy Sciences, Division of Materials Sciences and Engineering under award DE-SC0021984 through Stanford University (Z.Z., M.H., A.J.M.) and under FWP 100740 through SLAC National Accelerator Laboratory (Z.Z., M.H., A.J.M., T.F.H., D.G.-G.). Additional funding for fabrication and measurement of transistors was provided by SUPREME Center, jointly sponsored by the SRC and DARPA, from TSMC under the Stanford SystemX Alliance (L.H., E.P.), and from the Precourt Institute for Energy at Stanford University. This work was completed in part at the Stanford Nano Shared Facilities (SNF), supported by the National Science Foundation under award ECCS-2026822, and at the nano@Stanford labs, supported by the National Science Foundation as part of the National Nanotechnology Coordinated Infrastructure award ECCS-1542152. M.H. acknowledges partial support from the Department of Defense through the Graduate Fellowship in STEM Diversity program. J.H. acknowledges partial support from an NTT Graduate Research Fellowship. G.Z.Jr. and P.R. acknowledge support from the National Science Foundation Graduate Research Fellowship and Stanford Graduate Fellowship in Science & Engineering. The authors thank Anh Tuan Hoang and Kunal Mukherjee for the helpful discussions and comments on the manuscript.

**For Table of Contents Only**

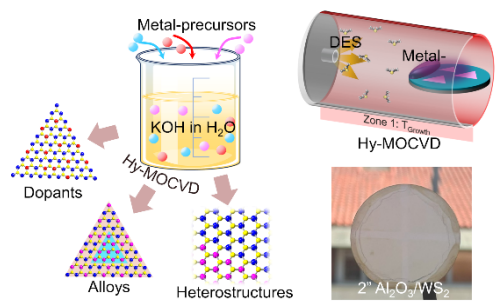